\documentclass[journal]{IEEEtran}
\IEEEoverridecommandlockouts 
\usepackage{cite}
\usepackage{amsmath,amssymb,amsfonts}
\usepackage{graphicx}
\usepackage{textcomp}
\usepackage{xcolor,soul,framed} 
\usepackage{mdwtab}
\usepackage{eqparbox}
\usepackage{array}
\usepackage[caption=false,font=normalsize,labelfont=sf,textfont=sf]{subfig}
\usepackage{textcomp}
\usepackage{array}
\usepackage{stfloats}
\usepackage{url}
\usepackage{verbatim}
\usepackage{cite}
\usepackage{multirow}
\usepackage{indentfirst}
\usepackage{booktabs}
\usepackage{breqn}
\usepackage{bm}
\usepackage{makecell}

\def\BibTeX{{\rm B\kern-.05em{\sc i\kern-.025em b}\kern-.08em
T\kern-.1667em\lower.7ex\hbox{E}\kern-.125emX}}
\usepackage[ruled]{algorithm2e}
\usepackage{algpseudocode}
\usepackage{url}
\usepackage{hyperref} 
\usepackage{xcolor}

\hypersetup{
    colorlinks=true,
    linkcolor=blue, 
    filecolor=black,
    urlcolor=blue, 
    citecolor=blue, 
    linkbordercolor={1 1 1}, 
    pdfborder={0 0 0} 
}
\graphicspath{ {./plots/} }
\begin{document}

 \title{Electric Arc Furnaces Scheduling under Electricity Price Volatility with Reinforcement Learning
}

\author{
\IEEEauthorblockN{Ruonan Pi, Zhiyuan Fan, Bolun Xu, \textit{Member, IEEE}  
	\thanks{ The authors are with Earth and Environmental Engineering, Columbia University, New York, NY 10027, USA. Email:\{rp3255, zf2198, bx2177\}@columbia.edu.}
}}

\maketitle

\begin{abstract}
This paper proposes a reinforcement learning-based framework for optimizing the operation of electric arc furnaces (EAFs) under volatile electricity prices. We formulate the deterministic version of the EAF scheduling problem into a mixed-integer linear programming (MILP) formulation, and then develop a Q-learning algorithm to perform real-time control of multiple EAF units under real-time price volatility and shared feeding capacity constraints. We design a custom reward function for the Q-learning algorithm to smooth the start-up penalties of the EAFs. Using real data from EAF designs and electricity prices in New York State, we benchmark our algorithm against a baseline rule-based controller and a MILP benchmark, assuming perfect price forecasts. The results show that our reinforcement learning algorithm achieves around 90\% of the profit compared to the perfect MILP benchmark in various single-unit and multi-unit cases under a non-anticipatory control setting.



\end{abstract}


\section{Introduction}
Steel production accounts for approximately 7-9\% of global greenhouse gas emissions and is widely considered one of the hardest sectors to decarbonize~\cite{kim_decarbonizing_2022}. Among decarbonization pathways, the electric arc furnace (EAF) is increasingly recognized as a critical technology, as it primarily utilizes scrap steel and electricity instead of iron ore and coke~\cite{fan_low-carbon_2021}. This shift substantially reduces direct emissions and enables flexible interaction with power systems, because the process is entirely electricity-driven. In particular, EAFs have the potential to shift production in response to volatile electricity prices and variable renewable generation, creating opportunities for both cost reduction and renewable integration.

Furnace operation involves non-trivial start-up costs, production delays, and feedstock limitations, resulting in a complex multi-period scheduling problem. The process embeds non-convex start-up, dwell-time, and sequencing logic, and at fine time resolution over long horizons, with multiple parallel furnaces, even tractable surrogates can be computationally burdensome for real-time use. Prior work therefore mostly adopts continuous-time Resource-Task Network and general-precedence formulations for EAF melt-shop scheduling under time-varying electricity prices\cite{zhang_cost-effective_2017}, scalable continuous variants \cite{lyu_efficient_2025}, and market or incentive mechanisms that shape the behavior of industrial loads \cite{mays_quasi-stochastic_2021,zheng_energy_2023,qi_locational_2025,li_socially_2024}. There is also a work embeds these models into energy-aware mixed-integer linear programming (MILP) frameworks that extend continuous-time schedules with explicit optimization of daily electricity purchases and sales under multiple contracts \cite{hadera_optimization_2015}. 



Practical deployment is further hindered by uncertainty in electricity prices, as day-ahead forecasts only imperfectly anticipate real-time spikes and structural shifts, so operators must act on noisy, biased price signals rather than the ex-post prices assumed in most models.
Incorporating this uncertainty into plant-scale MILPs is not trivial because stochastic or robust formulations require large scenario sets and additional binaries, which quickly become computationally prohibitive at 5-min resolution and across multiple furnaces. As a result, most plant-scale studies assume perfect price foresight and solve a single deterministic problem, often for one furnace at a time, leaving limited evidence on multi-furnace coordination or on how day-ahead and real-time decisions should interact under realistic forecast errors. 

We address these gaps by coupling a physics-based, rolling-horizon MILP
with continuous-progress variables in the continuous Resource--Task Network (cRTN) style,
and a Q-learning policy trained on day-ahead price signals. The MILP provides interpretable, upper-bound benchmarks under perfect foresight, while the RL dispatcher operates directly on price features without relying on commercial optimization solvers, making it attractive for industrial deployment within existing control stacks. Our framework coordinates three homogeneous EAFs under feeder limits using NYISO prices, delivering both feasible, stability-aware schedules and adaptive real-time performance. This paper makes the following contributions:
\begin{itemize}
\item \textbf{Methodology:} We develop a rolling-horizon MILP model that captures furnace start-up costs, production delays, and global resource constraints, and benchmark it against a solver-free Q-learning approach that learns dispatch policies from historical day-ahead price data.
\item \textbf{Modeling:} We extend from single-furnace to multi-furnace scheduling by introducing coupling constraints on aggregate power consumption and feedstock usage, thereby providing a closer representation of industrial practice.
\item \textbf{Numerical Analysis:} Using one year of NYISO West zone data, including both real-time and day-ahead prices. We quantify the trade-offs between optimization-based and learning-based approaches in terms of profitability, computational burden, and operational flexibility, and highlight the scalability of the RL policy for industry applications.
\end{itemize}

The remainder of the paper is organized as follows. Section~\ref{formulation} introduces the problem formulation. Section~\ref{method} describes the MILP and Q-learning approaches. Section~\ref{conclusion} concludes with insights and future research directions.

\section{Background and Related Work}
\label{sec:related}

\subsection{EAF Modeling and Scheduling}
Electric arc furnaces (EAFs) are central to scrap-based steelmaking and, due to large and adjustable power demand, are highly exposed to time-varying electricity prices. As a result, EAF operation is naturally formulated as a mixed-integer scheduling problem that must capture stage logic and power constraints. 
Early studies developed energy-aware MILP formulations that co-optimize production sequences with electricity procurement under time-sensitive tariffs \cite{hadera_optimization_2015}. 
For EAF-specific applications, resource--task network (RTN) models encode melting stages and transformer-tap flexibility, leveraging demand-side flexibility to reduce costs \cite{zhang_cost-effective_2017}. 
Complementing schedule-level models, process-oriented optimization and simulation work characterize the fidelity--tractability trade-off for online use and motivate reduced-order, control-oriented representations \cite{abadi_review_2024,saboohi_optimization_2019}.

To improve scalability, Lyu \textit{et al.} reformulate legacy RTN as a cRTN that replaces binary execution decisions with continuous task-progress variables and unifies resource-balance, execution, waiting-time, and production-target constraints; reported results show order-of-magnitude speedups while preserving modeling accuracy \cite{lyu_efficient_2025}. 
Such continuous surrogates are particularly attractive when frequent re-optimization is required (e.g., under high-frequency price signals or multi-furnace coordination).

Beyond the plant scale, system-level studies incorporate EAFs as flexible loads to support renewable integration. 
Zhao \textit{et al.} propose a two-stage (day-ahead and intra-day) scheduling framework that couples EAF demand response with wind-power modal decomposition and co-optimizes BESS and thermal units; in a real case, including EAF loads reduces wind curtailment by 40.49\% and day-ahead CO$_2$ emissions by 2.5\%, solved via an iterative genetic algorithm with CPLEX \cite{zhao_two-stage_2024}. 
These results reinforce the role of large, price-responsive industrial loads in system-level flexibility, complementing plant-scale MILP/RTN and continuous formulations \cite{hadera_optimization_2015,zhang_cost-effective_2017,lyu_efficient_2025}.

In practice, melt shops typically operate a \emph{limited number} of parallel major units rather than large fleets; public sources document sites with \emph{three} EAFs in operation \cite{noauthor_steel_nodate,noauthor_steelmaking_nodate}. 
Consistent with this few-unit layout and to keep problem size tractable, many plant-level formulations adopt parallel \emph{identical} (homogeneous) units for steelmaking stages, including the EAF stage \cite{zhang_cost-effective_2017,li_constraint_2015,su_multi-objective_2023}.

\subsection{Learning-based Industry Process Optimization}
Electricity tariffs and market designs materially shape optimal EAF operation. At the plant level, treating EAFs as controllable loads in multi-energy MILPs yields significant cost reductions under energy and demand charges \cite{lee_optimization_2023}. 
At the market level, quasi-stochastic clearing improves the handling of uncertainty in deterministic formulations, and state-of-charge (SoC) segment models and locational bid bounds align flexible-resource bids with social welfare under non-convexities and risk \cite{mays_quasi-stochastic_2021,zheng_energy_2023,qi_locational_2025}. 
From the incentive side, adaptive two-time-scale pricing achieves socially optimal consumption without requiring disclosure of user-private models, indicating the feasibility of learning-compatible coordination \cite{li_socially_2024}. 
Relatedly, learning and optimization have been coupled in grid applications, such as threshold policies for regulation and chance-constrained, to achieve adaptive performance under uncertainty \cite{xu_optimal_2018}.
In practice, however, operators must satisfy start-up, dwell-time, and feeder limits while responding to nonstationary prices without proprietary foresight. We therefore combine a rolling-horizon MILP for guaranteed feasibility with a learned policy for low-latency adaptation.

\section{Problem Formulation}\label{formulation}

We consider the short-term operation of an electric arc furnace (EAF) system participating in a wholesale electricity market. The system consists of one or multiple furnaces under the same operator, where electricity is the dominant input and steel output is subject to furnace physics and operational constraints. This formulation establishes a deterministic baseline against which more advanced scheduling strategies (e.g., rolling-horizon or reinforcement learning) can be benchmarked.  
\begin{table}[!ht]
\renewcommand{\arraystretch}{1.2}
\caption{Nomenclature for the EAF scheduling model}
\label{Nomenclature}
\centering
\setlength{\tabcolsep}{5mm}{%
\begin{tabular}{c c}
\toprule
\textbf{Symbol} & \textbf{Description} \\
\midrule
\multicolumn{2}{l}{\textbf{Sets and indices}} \\
$i \in \mathcal{N}$ & Index of furnaces \\
$t \in \mathcal{T}$ & Index of time periods (5-min steps) \\
\midrule
\multicolumn{2}{l}{\textbf{Decision variables}} \\
$i_{i,t}$ & Material input [ton] \\
$r_{i,t}$ & Production (tapping) rate [ton/step] \\
$k_{i,t}$ & Melting rate [ton/step] \\
$m_{i,t}$ & Molten stock [ton] \\
$s_{i,t}$ & Solid stock [ton] \\
$u_{i,t}$ & Binary: furnace on/off \\
$v_{i,t}$ & Binary: melting-stage indicator \\
$y_{i,t}$ & Binary: startup indicator \\
$P_{i,t}$ & Power consumption [MW] \\
\midrule
\multicolumn{2}{l}{\textbf{Parameters}} \\
$\lambda_t$ & Market clearing price [\$/MWh] \\
$P^{\text{base}}_i$ & Base load [MW] \\
$P^{\text{melt}}_i$ & Extra load during melting [MW] \\
$R_i$ & Max production rate [ton/step] \\
$K_i$ & Max melting rate [ton/step] \\
$I_i$ & Batch input size [ton] \\
$P^{\max}$ & System-wide power cap [MW] \\
$\pi$ & Market price of production [\$/ton] \\
$\alpha_i$ & Material-to-product conversion ratio \\
$C$ & Processing cost [\$/ton] \\
$\delta_i$ & Startup penalty [\$] \\
\bottomrule
\end{tabular}%
}
\end{table}

\subsection{Electricity Market Model}
The system interacts with wholesale electricity markets through locational marginal prices (LMPs) $\lambda_t$ (\$/MWh). The planning horizon is discretized into $\mathcal{T} = \{1,2,\dots,T\}$ with an interval $\Delta t=5$ minutes.  
Within each time interval $t$, the blast furnace's electricity expenditure $\lambda_t P_{i,t}$ is proportional to its electricity consumption $P_{i,t}$ (unless the spot electricity price is negative at that moment), while the revenue from iron production is proportional to the production rate $r_{i,t}$.  
In this baseline formulation, real-time prices are assumed to be known in advance, providing a perfect-foresight benchmark that serves as an upper bound on achievable profit.  

\subsection{Furnace Operation Model}

\begin{figure}[!ht]
\setlength{\abovecaptionskip}{-0.1cm}  
    \setlength{\belowcaptionskip}{-0.1cm} 
    \centering
    \includegraphics[width=\columnwidth]{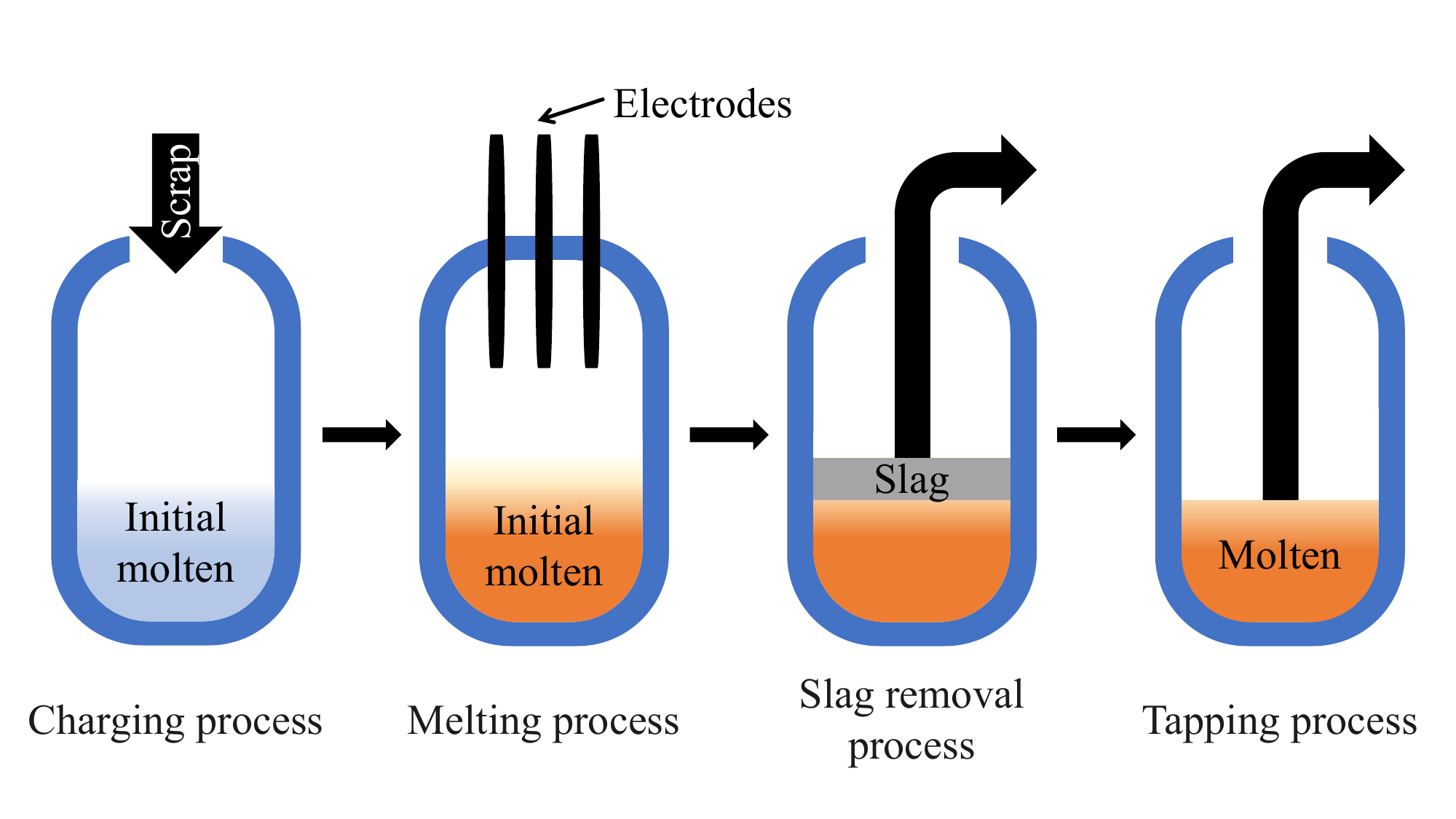}
    \caption{Electric arc furnace (EAF) batch cycle. The physical process consists of
    charging, melting, slag removal, and tapping. For optimization, we aggregate these
    into a high-power melting stage and a base-power stage (charging/slag/tapping).}
    \label{fig:eaf-cycle}
\end{figure}

The electric arc furnace (EAF) operates in repetitive batch cycles (Fig.~\ref{fig:eaf-cycle}).
A typical cycle includes four steps: \emph{charging} (scrap/DRI is loaded on top of an
initial molten heel), \emph{melting} (arcs supply high electrical power to melt the solid
charge), \emph{slag removal} (oxidation by-products are skimmed), and \emph{tapping}
(Liquid steel is discharged to the ladle.) Electrical demand is strongly stage-dependent:
the melting step dominates energy use, while the other steps require substantially lower
power for holding, mixing, and handling.

For tractability, we abstract the cycle into two characteristic power stages:
(i) a \textbf{high-power} stage representing melting, and (ii) a \textbf{base-power} stage
representing charging/slag/tapping. This preserves the key flexibility of EAFs (alternation
between energy-intensive and low-load periods) while enabling a compact optimization model.

All variables, parameters, the objective function, and constraints are formally defined 
in Table~\ref{Nomenclature}.
Let $t\in\mathcal{T}$ index discrete time periods. Each furnace $i \in \mathcal{N}$ transitions among operational stages governed by binary variables.

Decision binary variables are:
$u_{i,t}\in\{0,1\}$ (furnace on/off), $v_{i,t}\in\{0,1\}$ (high-power/melting indicator),
$y_{i,t}\in\{0,1\}$ (startup indicator).

The profit of an EAF operator can be represented as market revenue minus operating costs. 
The revenue term $\pi \alpha_i r_{i,t}$ reflects that sales are only realized when steel is tapped: the effective output is given by the tapping rate $r_{i,t}$ multiplied by the yield coefficient $\alpha_i$, and is then valued at the market price $\pi$. 
On the cost side, $C i_{i,t}$ accounts for the processing of raw material associated with each batch input, while $\lambda_t P_{i,t}$ captures the electrical expenditures, where the power draw of the furnace is settled at the prevailing market clearing price. 
Finally, a startup penalty $\delta_i$ is added whenever $y_{i,t}=1$, representing additional wear of the electrodes, thermal stress, and auxiliary resource usage whenever a new batch is initiated. Since the producer ultimately seeks to maximize profit, these revenue and cost components together form the objective function in~\eqref{obj}.

\begin{figure}[t]
\centering
\includegraphics[width=\linewidth]{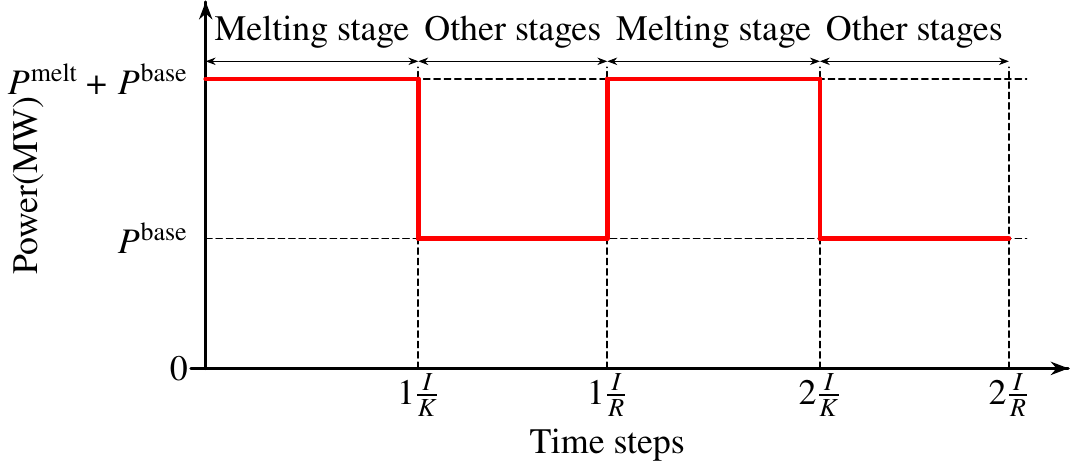}
\caption{Simplified two-stage power profile for an EAF cycle.High-power (melting) alternates with base-power (charging/slag/tapping).}
\label{fig:two-stage-profile}
\end{figure}

We use nameplate parameters $P^{\mathrm{base}}_i$ and $P^{\mathrm{melt}}_i$ to denote the baseline power consumption when furnace $i$ is operating, and the additional power required during its melting stage.
A two-stage power model is discribed by \eqref{power} and \eqref{startup_logic}, which enforces: (a) melting(high-power) only when the furnace is on; (b) startup detection;
and (c) a piecewise-cap on $P_t$ that yields $P_t = P^{\mathrm{base}}$ in the base-power stage and $P_t=P^{\mathrm{base}} + P^{\mathrm{melt}}$ in the high-power stage(Fig.~\ref{fig:two-stage-profile}).

\begin{figure}[t]
\centering
\includegraphics[width=\linewidth]{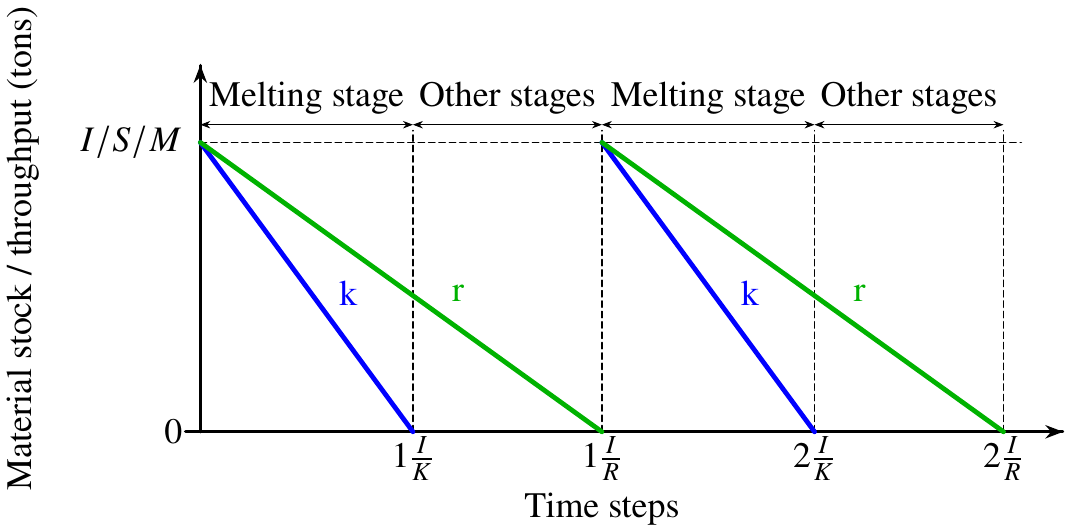}
\caption{Two-stage abstraction of the EAF cycle. 
\textbf Schematic material dynamics over time. 
The melting throughput $k$ (blue) is activated during the high-power stage, while 
the production/tapping throughput $r$ (green) is scheduled when the furnace is on. 
Dashed vertical lines mark stage-switching instants; axis annotations indicate nominal durations/scales.} 
\label{fig:krms}
\end{figure}

We distinguish two rate variables in the furnace cycle: 
$r_{i,t}$, the \emph{production (tapping) rate}, and 
$k_{i,t}$, the \emph{melting rate} (Fig.~\ref{fig:krms}). 
The nameplate parameters $R_i$ and $K_i$ denote the maximum feasible rates for 
production and melting, respectively. 
Their realizations are tied to the operating stage indicators $u_{i,t}$ (furnace on/off) 
and $v_{i,t}$ (melting/high-power) by \eqref{rate_caps_r} and \eqref{rate_caps_k}.

To ensure feasibility, the actual tapping and melting rates cannot exceed the available material stocks, we have \eqref{r_material_limit} and \eqref{k_material_limit}, where $s_{i,t}$ and $m_{i,t}$ track the solid-charge stock and molten stock, respectively.

Together, \eqref{input_startup} and the inventory dynamics follow balance constraints \eqref{state_updates_m}--\eqref{nonneg_ms} enforce batch charging and material conservation: the molten stock $m_{i,t}$ accumulates newly fed input $i_{i,t}$ and decreases with melting, while the solid stock $s_{i,t}$ evolves analogously with tapping. This abstraction ensures that production can only occur if sufficient stock is present and that melting/tapping are aligned with the furnace operating stages depicted in Fig.~\ref{fig:two-stage-profile} and Fig.~\ref{fig:krms}.

In addition to individual furnace dynamics, system-level constraints impose aggregate limits across all units. Specifically, the total instantaneous power demand
is capped by \eqref{totalpower_cons}.
 
Constraints~\eqref{cont_vars} define the continuous decision variables (material flows and power consumption) as nonnegative real numbers. 
Finally,~\eqref{init_start} and ~\eqref{binary_start} initialize the system at $t=0$ with empty inventories and no active operation, ensuring that the first batch must be explicitly started.

\subsection{Optimization Problem}
As the operator seeks to maximize profit defined as steel revenue net of material, electricity, and startup costs, combining the above,the deterministic baseline problem can be written as the mixed-integer linear program (MILP) in~\eqref{EAF}, for all $i \in \mathcal{N}$ and $t \in \mathcal{T}$.


\begin{subequations}\label{EAF}
\begin{align}
&\begin{aligned}
\max_{\mathbf{x}} \sum_{t\in\mathcal{T}}\sum_{i\in\mathcal{N}} \Big(
   \pi \alpha_i r_{i,t} - C i_{i,t} - \lambda_t P_{i,t} - \delta_i y_{i,t} \Big)
\end{aligned}\label{obj}\\[4pt]
\text{s.t. }\;
% ...
& P_{i,t} = P^{\mathrm{base}}_i u_{i,t} + P^{\mathrm{melt}}_i v_{i,t}, && \label{power}\\
& y_{i,t} \le 1 - \tfrac{u_{i,t-1}+v_{i,t-1}}{2}, &&\label{startup_logic}\\
& 0 \le r_{i,t} \le R_i u_{i,t}, && \label{rate_caps_r}\\
& 0 \le k_{i,t} \le K_i v_{i,t}, && \label{rate_caps_k}\\
& r_{i,t} \le s_{i,t}, && \label{r_material_limit}\\
& k_{i,t} \le m_{i,t}, &&\label{k_material_limit}\\
& i_{i,t} = I_i y_{i,t}, &&\label{input_startup}\\
& m_{i,t} = m_{i,t-1} + i_{i,t} - k_{i,t-1}, &&\label{state_updates_m}\\
& s_{i,t} = s_{i,t-1} + i_{i,t} - r_{i,t-1}, && \label{state_updates_s}\\
& m_{i,t}, s_{i,t} \ge 0, &&\label{nonneg_ms}\\
& \sum_{i\in\mathcal{N}} P_{i,t} \le P^{\max},&& \label{totalpower_cons}\\
& u_{i,t}, v_{i,t}, y_{i,t} \in \{0,1\}, && \label{bin_uvy}\\
& r_{i,t}, k_{i,t}, i_{i,t}, P_{i,t} \in \mathbb{R}_+, &&\label{cont_vars}\\
& m_{i,0}=s_{i,0}=r_{i,0}=k_{i,0}=0, && \label{init_start}\\
& u_{i,0}=v_{i,0}=0, &&\label{binary_start}
\end{align}
\end{subequations}

\noindent with decision vector.
\[
\mathbf{x} =
\{ i_{i,t}, r_{i,t}, k_{i,t}, m_{i,t}, s_{i,t},
u_{i,t}, v_{i,t}, y_{i,t}, P_{i,t} \}_{i\in\mathcal{N},\, t\in\mathcal{T}}.
\]
The problem~\eqref{EAF} defines a profit-maximizing benchmark under perfect foresight of real-time prices. 
In practice, however, solving this MILP at a five-minute resolution for a full year is computationally prohibitive, and the perfect-foresight assumption does not hold. 
To address these issues, we adopt a rolling-horizon approximation and later introduce a reinforcement learning framework described in Section~\ref{method}.
However, the MILP benchmark remains useful as a deterministic reference to evaluate alternative scheduling strategies.

\section{Methodology}\label{method}

\subsection{Rolling-horizon MILP}

\begin{algorithm}[htbp]\label{alg:rh-milp}
\hspace*{\fill}
\caption{Rolling-Horizon MILP Scheduling Procedure}
\SetAlgoLined
\SetAlgoNoEnd
\KwIn{Real-time price series $\{\lambda_t\}_{t=1}^T$, furnace configs $\{I_i,R_i,K_i,\pi\alpha_i,C_i,\delta_i,P^{\text{base}}_i,P^{\text{melt}}_i\}_{i\in\mathcal N}$, system caps $P_{\max},I_{\max}$, horizon $H$, step size $S$.}
\KwOut{Dispatch decisions $x=\{i_{i,t},r_{i,t},k_{i,t},m_{i,t},s_{i,t},u_{i,t},v_{i,t},y_{i,t},P_{i,t}\}$.}
\SetKwBlock{StepOne}{Algorithm:}{}
\StepOne{
    Initialize states $(m_i,s_i,u_i,v_i,\text{last-}u_i,\text{last-}v_i,\text{last-}r_i,\text{last-}k_i)$ for all $i\in\mathcal N$, cumulative profit $\Pi=0$ \\
    \While{$t \leq T$}{
        Set window $W = \{t,\dots,\min(t+H-1,T)\}$ \\
        Build MILP with objective
        $\max \sum_{\tau \in W}\sum_{i\in\mathcal N} \big(\pi\alpha_i r_{i,\tau}-C_i r_{i,\tau}-\lambda_\tau P_{i,\tau}-\delta_i y_{i,\tau}\big)$ \\
        subject to constraints (1b)-(1o), aggregate caps $\sum_i P_{i,\tau}\le P_{\max}$, $\sum_i i_{i,\tau}\le I_{\max}$, and boundary conditions \\
        Solve MILP over $W$ using GUROBI \\
        \If{optimal solution found}{
            Implement first $S$ steps: update dispatch, log profit, $\Pi \gets \Pi + \sum_{t}^{t+S-1}\text{profit}$ \\
            Carry terminal states $(m,s,u,v,r,k)$ to next window
        }
        \Else{
            Save checkpoint; optionally relax gap or skip window
        }
        $t \gets t+S$
    }
    Return cumulative profit $\Pi$ and full dispatch trajectory
}
\end{algorithm}

At each iteration $k$, the horizon is restricted to 
$W_k=\{t_k,\dots,t_k+H-1\}$, where $H$ is the look-ahead horizon 
and $S$ is the step size. 
We solve the MILP over $W_k$, apply only the first $S$ decisions, 
and then roll forward to $t_{k+1}=t_k+S$.

\subsubsection{Constraints within the window} 
The window problem inherits all furnace-level constraints 
from Problem~\eqref{EAF}, including rate caps, inventory balances, 
and power definitions, but restricted to $t\in W_k$. 
In addition, boundary conditions are introduced to link the new window 
to the terminal states of the previous one:
\begin{align}
m_{i,t_k} &= \bar m_i^{(k)} + i_{i,t_k} - \bar k_i^{(k)}, \label{eq:rh_m}\\
s_{i,t_k} &= \bar s_i^{(k)} + i_{i,t_k} - \bar r_i^{(k)}, \label{eq:rh_s}\\
y_{i,t_k} &\le 1 - \tfrac{\bar u_i^{(k)} + \bar v_i^{(k)}}{2}, \label{eq:rh_y}
\end{align}
where $\bar m_i^{(k)}, \bar s_i^{(k)}, \bar u_i^{(k)}, \bar v_i^{(k)}, 
\bar r_i^{(k)}, \bar k_i^{(k)}$ are the terminal states carried over 
from the previous window. 
System-wide limits are enforced for each period $\tau\in W_k$:

\begin{align}
\sum_{i\in\mathcal N} P_{i,\tau} \le P_{\max}
\end{align}

\subsubsection{State update} 
After implementing the first $S$ steps, terminal states are updated as
\begin{align}
 m_i^{(k+1)} &= m_{i,t_k+S},\\
 s_i^{(k+1)} &= s_{i,t_k+S},\\
 u_i^{(k+1)} &= u_{i,t_k+S-1},\\
 v_i^{(k+1)} &= v_{i,t_k+S-1},
\end{align}
with analogous updates for $\bar r_i^{(k+1)}$ and $\bar k_i^{(k+1)}$.
These carried-over variables serve as the initial conditions for the next window, so that material inventories and startup states evolve consistently over time. 

In this setting, the rolling-horizon MILP can be viewed as a practical relaxation of the full clairvoyant benchmark: a larger horizon $H$ allows the solution to approximate the year-ahead optimum more closely, but increases computational burden, while a smaller step size $S$ forces more frequent re-optimization. 
This trade-off between fidelity and tractability provides the baseline against which we later compare the reinforcement learning approach.

Although this rolling-horizon MILP offers a tractable relaxation of the full-year problem, it still relies on perfect knowledge of future real-time prices, which is not available in the real world.

\subsection{Q-learning Framework}

To overcome the unrealistic perfect-foresight assumption of the rolling-horizon MILP,
we develop a reinforcement learning (RL) framework that learns adaptive dispatch policies 
using only day-ahead market information.
A tabular Q-learning agent interacts with a simplified multi-unit EAF environment,
observes local furnace states and day-ahead (DAP) price signals,
and gradually learns profitable operation strategies from historical data.  
This learning-based formulation enables implementable, data-driven decision making
without requiring future real-time price trajectories.

\subsubsection{State Representation}

At each time step $t$, the environment state includes the operational status of all furnaces
and coarse-grained market information available from the day-ahead market:
\[
s_t = \Big[z_t,\, \{\tau_{i,t}\}_{i\in\mathcal{N}}\Big].
\]
Here $\tau_{i,t}\!\in\!\{0,\dots,L\}$ denotes the remaining steps of furnace $i$’s current cycle
($L\!=\!16$ in our implementation, including $12$ melting and $4$ tapping steps),
and $z_t$ is a discrete DAP price bucket derived from the average and short-term trend of
day-ahead prices.
This state captures both operational memory (through $\tau_{i,t}$) and limited forward
market expectation (through $z_t$), approximating the information set realistically available
to the operator.

\subsubsection{Action Space and Feasibility Masking}

The agent chooses a joint startup action
\[a_t \in \{0,1\}^{|\mathcal{N}|},\]
where $a_{i,t}=1$ triggers furnace $i$ to start if currently idle.
The joint action must satisfy global operational limits:
\[
\sum_{i\in\mathcal{N}} P_{i,t}(a_t,\bm{\tau}_t) \le P^{\max}
\]
Infeasible actions violating these limits are masked before selection.
An $\epsilon$-greedy strategy with a slowly decaying $\epsilon$
balances exploration and exploitation during training.

\subsubsection{Reward Design with Adjustable Penalty Smoothing}

The instantaneous reward represents the total operating profit of all units:
\begin{equation}\label{eq:reward}
r_t = \sum_{i\in\mathcal{N}}
  \Big(\pi \alpha_i r_{i,t} - C r_{i,t}
       - \lambda^{\mathrm{DAP}}_t P_{i,t}
       - \tfrac{\delta_i}{\kappa_i}\,{active}_{i,t}\Big),
\end{equation}
where ${active}_{i,t}\!\in\!\{0,1\}$ indicates that the furnace $i$ is in the melting or tapping stage.
Instead of charging the entire startup cost $\delta_i$ at once,
we evenly distribute it across $\kappa_i$ active steps to smooth reward fluctuations
and stabilize the temporal-difference updates.
Importantly, $\kappa_i$ serves as a \textit{tunable shaping coefficient}:
a larger $\kappa_i$ spreads the penalty thinner, encouraging more frequent startups
and higher throughput (aggressive scheduling),
whereas a smaller $\kappa_i$ concentrates the penalty, promoting conservative,
profit-oriented operation.
Hence, $\kappa_i$ can be viewed as a control knob that adjusts the 
“production aggressiveness” of the learned policy.

\subsubsection{Learning Algorithm and Heuristic Tie-breaking}

The agent maintains a tabular Q-value for each feasible state-action pair and updates via
\[
Q(s_t,a_t)\leftarrow (1-\eta)Q(s_t,a_t)
+\eta\Big[r_t+\gamma\max_{a'\in\mathcal{A}(s_{t+1})}Q(s_{t+1},a')\Big],
\]
with learning rate $\eta$ and discount factor $\gamma$.
When multiple actions yield identical Q-values, we apply a short-horizon 
template-based lookahead to break ties:
\[
\hat V^{\mathrm{LA}}(t,a) =
\sum_{i:\,a_i=1}\sum_{s=0}^{L-1}
\Big[(\pi\alpha_i-C)\bar r_i(s) - \tilde{\lambda}_{t+s}\bar P_i(s)\Big] - \delta_i,
\]
where $(\bar r_i(s), \bar P_i(s))$ follow the predefined melt/tap pattern of unit $i$.
This heuristic preserves the sequence structure of furnace cycles while maintaining
computational efficiency.

\subsubsection{Training and Evaluation Protocol}

Training is performed on 2023 trajectories using day-ahead price features only.
The learned policy is then tested on 2024 trajectories with $\epsilon=0$
to evaluate its out-of-sample performance.
During testing, actions are executed greedily based on the learned Q-table,
and realized profits are computed ex post using real-time (RTP) prices.
Performance metrics include cumulative profit, number of startups,
utilization rates, and daily profit variance, allowing comparison 
with the MILP benchmark introduced earlier.

\subsubsection{Discussion}

The proposed Q-learning framework provides a practical, information-limited counterpart
to the clairvoyant MILP benchmark.
It learns implementable dispatch policies using only day-ahead data,
and once trained, inference is instantaneous and suitable for real-time operation.
Reward smoothing with coefficient $\kappa_i$ not only stabilizes learning
but also serves as an interpretable economic control knob linking 
learning dynamics and production aggressiveness—
analogous to a temperature parameter in stochastic control.
This feature enables the operator to tune between yield-oriented and
profit-oriented scheduling behaviors within a unified RL formulation.

\subsection{Baseline Policies}

To contextualize the optimization and learning results, 
we implement several fixed-operation baselines that follow deterministic
production cycles without any price-dependent decision-making.
These baselines provide lower-bound references for evaluating 
the economic value of adaptive control.

\subsubsection{Fixed-Cycle Operation}

In the fixed-cycle benchmark, each furnace repeats its nominal
melting-tapping-cooling sequence regardless of market conditions.
For furnace $i$, the predefined cycle length is 
$L_i=L_i^{\mathrm{melt}}+L_i^{\mathrm{tap}}+L_i^{\mathrm{stop}}$,
where $L_i^{\mathrm{melt}}$ and $L_i^{\mathrm{tap}}$ denote the melting and tapping durations,
and $L_i^{\mathrm{stop}}$ represents an idle cooldown phase.
At every time step, the operating indicators $\{u_{i,t},v_{i,t}\}$ 
and startup flag $y_{i,t}$ are determined by the fixed cycle position.
No optimization or rescheduling occurs, and 
the startup cost $\delta_i$ is charged each time the unit re-enters
the melting phase.

The total system power is monitored under the same
limit as in the MILP and Q-learning settings:
\[
\sum_i P_{i,t} \le P^{\max}
\]
and because the limit, only two furnaces can operate concurrently.

We evaluate the scenario that operate 
two furnaces concurrently under fixed cycles:
The simulation spans one year of 5-minute data (2024 real-time prices),
yielding $105{,}120$ time steps.
Then, we record energy consumption, material throughput, 
startup frequency, and total profit:
\[
\Pi_{\mathrm{fixed}} = 
\sum_{t,i} (\pi \alpha_i r_{i,t} - C r_{i,t}
- \lambda^{\mathrm{RTP}}_t P_{i,t} - \delta_i y_{i,t}).
\]
Because furnaces run continuously and ignore price signals, these fixed-cycle cases represent a non-adaptive lower bound
on achievable profit and serve as intuitive baselines for assessing the performance gains from MILP optimization and Q-learning policies.

\section{Results}

\subsection{Test System Setting}

\subsubsection{Multi-unit configuration}
Our baseline model considers a melt shop featuring three \emph{homogeneous} EAF units operating in parallel. 
This configuration is selected to balance industrial realism with computational tractability.
In terms of shop-floor reality, while many mini-mills operate with one or two Ultra-High Power (UHP) furnaces\cite{stubbles_minimill_2009}, capacity expansions at major facilities—such as North Star BlueScope—have validated the operational necessity of three-EAF layouts \cite{noauthor_steelmaking_nodate,steel_alexandria_nodate}.
From a modeling perspective, treating these parallel units as \emph{identical} is a standard abstraction in the literature, widely adopted in RTN-based and constraint-programming formulations to maintain problem solvability \cite{zhang_cost-effective_2017,li_constraint_2015,su_multi-objective_2023}.
Therefore, we adopt three identical furnaces as a representative high-complexity benchmark, reserving heterogeneous parameter settings for the sensitivity analysis.

\subsubsection{System Configuration and Parameters}

The test system models a small-scale steelmaking facility with three heterogeneous
electric arc furnaces (EAFs), differing in efficiency, capacity, and startup cost.
All units follow the same operational template of melting-tapping-cooling cycles
but with distinct cycle durations and power ratings.
Table~\ref{tab:paramshomo} summarizes the unit-level parameters used across all simulations.

\begin{table}[h]
\centering
\caption{Homogeneous EAF unit parameters used in all simulations.}
\label{tab:paramshomo}
\setlength{\tabcolsep}{4pt} 
\renewcommand{\arraystretch}{1.2}
\begin{tabular}{lccccccccc}
\toprule
\multirow{2}{*}{\textbf{Furnace}} & \multirow{2}{*}{$I_i$} & \multirow{2}{*}{$R_i$} & \multirow{2}{*}{$K_i$} & \multirow{2}{*}{$\pi_i$} & \multirow{2}{*}{$\alpha_i$} & \multirow{2}{*}{$C_i$} & \multirow{2}{*}{$\delta_i$} & \multicolumn{2}{c}{Power [MW]} \\ 
\cmidrule(lr){9-10}
 & & & & & & & & $P^{\text{melt}}$ & $P^{\text{base}}$ \\ 
\midrule
Standard & 1.0 & 1/12 & 1/15 & 400 & 0.92 & 300 & 50 & 0.0367 & 0.0033 \\
\bottomrule
\end{tabular}
\end{table}

All parameters are expressed per 5-minute step.
Each furnace $i$ requires $I_i$ units of feedstock to start a cycle,
produces molten steel at rate $R_i$ during the melting phase
and slag output at rate $K_i$ during tapping.
$\pi_i$ and $C_i$ represent unit selling price and processing cost per unit output,
$\alpha_i$ the yield factor, $\delta_i$ the startup cost,
and $(P^{melt},P^{base})$ the active and background power consumptions. Crucially, the power parameters are calibrated such that the total energy input per production cycle approximates 0.49~MWh/ton (derived from $P^{\text{melt}} \cdot  I_i/R_i + P^{\text{base}} \cdot I_i/K_i$). 
This value aligns closely with the U.S. industry average of approximately 440~kWh (0.44~MWh) of electricity per ton of crude steel reported in recent environmental assessments~\cite{pistorius_relating_2018}.
The global limit on simultaneous operation is$
P^{\max} = 0.107~\mathrm{MW}$ corresponding to the combined full-load capacity of the two largest furnaces.

\subsubsection{Market Data}

Real electricity price data are obtained from the 
New York Independent System Operator (NYISO) Western Zone
for both real-time and day-ahead markets.
Each dataset covers one full year (2023--2024) at a 5-minute resolution,
yielding 105{,}120 time steps per year.
Day-ahead (DAP) prices are used as inputs for Q-learning training and inference,
while real-time (RTP) prices are used to compute realized profits
for both MILP and RL evaluation.

\subsubsection{Simulation Environment}

All simulations are implemented in \texttt{Python~3.10} and executed on a workstation 
with an AMD Ryzen~9 5900X CPU (12~cores, 64~GB RAM) running \texttt{Windows~10}.
The optimization benchmark uses \texttt{CVXPY} with the \texttt{Gurobi~10.0} solver,
and the Q-learning agent is implemented in pure Python using 
\texttt{NumPy} and \texttt{tqdm} for efficient vectorized updates.
Each annual MILP trajectory (rolling-horizon optimization with 4-hour look-ahead)
takes approximately 3--5~minutes per window to solve,
while Q-learning training over 600 episodes requires around 20~minutes in total.
Fixed baselines are simulated over the same time horizon for direct comparison.

\subsubsection{Compared Settings}

Three modeling settings are evaluated under identical system and data inputs:

\begin{itemize}
  \item \textbf{Fixed-cycle baselines:} each furnace follows its nominal
  melting-tapping-cooling schedule without reacting to market prices,
  representing non-adaptive operation.

  \item \textbf{Q-learning policy:} a tabular RL agent learns 
  dispatch decisions using day-ahead prices and furnace states.
  Startup penalty smoothing coefficient $\kappa_i$
  adjusts the aggressiveness of the learned policy.

  \item \textbf{Rolling-horizon MILP benchmark:}
  an optimization-based clairvoyant solution assuming perfect foresight 
  of future real-time prices, representing an upper bound on achievable profit.
\end{itemize}

These three cases together establish a consistent benchmark suite:
the fixed-cycle runs provide the non-adaptive lower bound,
the Q-learning agent yields a feasible real-time policy under realistic information,
and the MILP benchmark quantifies the theoretical optimum under perfect foresight.



\subsection{Multi-unit Simulation with Coupled Grid Capacity}
\label{sec:multi-coupled}

\begin{figure}[!ht]
\setlength{\abovecaptionskip}{-0.1cm}  
    \setlength{\belowcaptionskip}{-0.1cm} 
    \centering
    \includegraphics[width=\columnwidth]{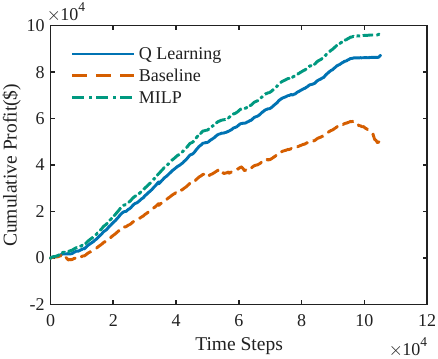}
\caption{Cumulative profit under a coupled feeder capacity $P_{\max}$ for MILP (clairvoyant), Q-learning (no foresight), and a heuristic baseline.}
    \label{fig:cumprofit}
\end{figure}

We evaluate three dispatch policies on a multi-furnace system under a coupled feeder capacity constraint $P_{\max}$ over a 60-day out-of-sample horizon with 5-minute steps.
The benchmark is a rolling-horizon MILP with perfect price foresight; the learning policy uses only information available at decision time; the baseline is a heuristic that ignores inter-unit coordination.
Figure~\ref{fig:cumprofit} reports cumulative profit.

\subsubsection{Headline results}
Over the full 60-day horizon, the MILP attains a cumulative profit of \$96{,}978, Q-learning reaches \$87{,}112, and the baseline achieves \$50{,}529.
Measured against the baseline, Q-learning delivers a \textbf{72.4\%} uplift in profit and attains \textbf{89.8\%} of the MILP benchmark (a \(\approx\)10.2\% gap to MILP).
These results indicate that the learned policy captures most of the clairvoyant upper bound while using only information available at decision time (Table~\ref{tab:profit_summary}).

\subsubsection{Effect of the coupled capacity}
Because all furnaces share \(P_{\max}\), simultaneous melting during high-price intervals induces a nonzero shadow price on the feeder.
The MILP staggers melt phases across units to keep aggregate load near, but not above, \(P_{\max}\) during peaks and backfills in troughs.
The learner reproduces much of this staggering: it (i) advances some melts into pre-peak ramps and (ii) defers some finishes into post-peak tails, which reduces curtailment at the cap and smooths net power.
In contrast, the baseline often stacks melters, hits the cap, and leaves profitable windows partially unexploited.

\begin{table}[t]
    \centering
    \caption{Profit over the 365-day horizon. Uplift is measured relative to the baseline; the last column reports the fraction of MILP profit achieved.}
    \label{tab:profit_summary}
    \setlength{\tabcolsep}{1.5pt}
    \begin{tabular}{lrrr}
        \toprule
        Policy & \multicolumn{1}{c}{Cumulative profit [\$]} & \multicolumn{1}{c}{Uplift vs Baseline} & \multicolumn{1}{c}{Share of MILP} \\
        \midrule
        Baseline    & 50{,}529  & ---       & 52.1\%  \\
        Q-learning  & 87{,}112  & +72.4\%   & 89.8\%  \\
        MILP        & 96{,}978  & +91.9\%   & 100\%   \\
        \bottomrule
    \end{tabular}
\end{table}

\begin{figure}[htbp]
    \centering
    \includegraphics[width=\linewidth]{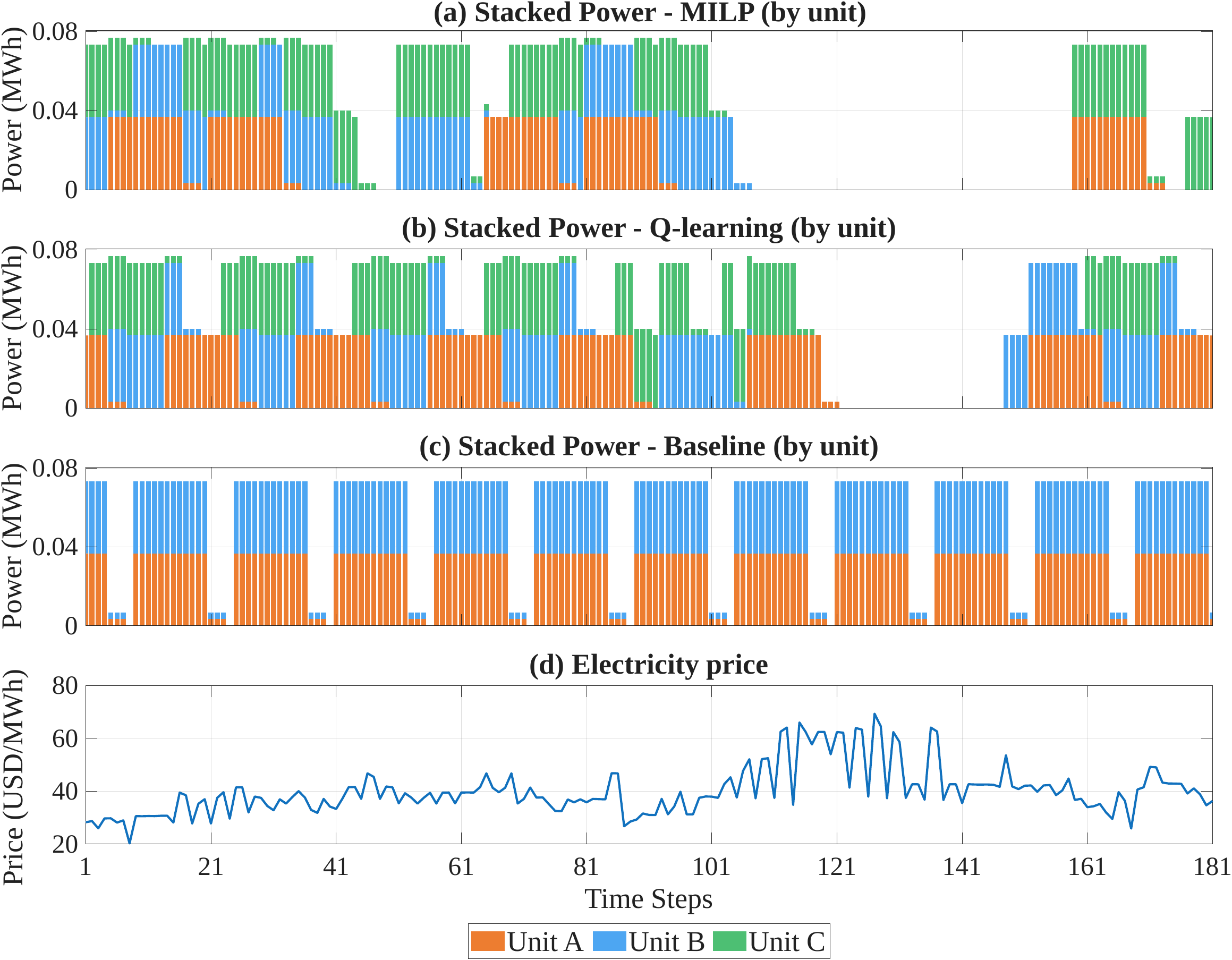}
    \caption{Comparison of per-unit furnace power trajectories under (a) MILP, (b) Q-learning, and (c) baseline scheduling over a representative 200$\times$5-min interval. Panels (a)--(c) show stacked power by unit, and panel (d) shows the corresponding electricity price. The MILP and Q-learning policies largely concentrate melting in low-price intervals and reduce load during price spikes, whereas the baseline follows a rigid, price-agnostic pattern that often maintains high power during expensive periods.}

    \label{fig:stacked_power}
\end{figure}

\subsubsection{Coordinated scheduling under dynamic prices}

Figure~\ref{fig:stacked_power} illustrates the dispatch patterns obtained from the MILP, Q-learning, and baseline approaches. 
Each bar represents one timestep, and different colors denote power consumption from individual furnaces.

Both the MILP and Q-learning frameworks exhibit clear coordination among units. 
During low-price intervals, multiple furnaces operate simultaneously to fully utilize the feeder capacity, 
while in high-price periods, production is reduced or paused to avoid uneconomic operation. 
This behavior effectively “fills the valleys” of the price profile — exploiting cheap electricity to maximize throughput — 
and “shaves the peaks” when energy costs surge.

In contrast, the baseline schedule keeps each furnace on a fixed duty cycle regardless of market signals, 
resulting in inefficient energy usage and reduced overall profit. 
The coordinated flexibility of MILP and Q-learning thus enables near-optimal load staggering across units, 
mimicking human-like scheduling decisions in response to electricity price fluctuations.

\subsubsection{Computation time}
Table~\ref{tab:runtime} compares the computational cost among all methods. 
The baseline rule-based dispatch completes within one minute on the full-year horizon (105{,}120 steps). 
MILP, which repeatedly solves mixed-integer programs for each receding window, 
requires about 1.47\,h in total (50\,ms per step on average). 
Q-learning involves a one-time offline training phase of 2.46\,h 
followed by a lightweight online inference of only 9.52\,s 
(0.09\,ms per step), achieving over \textbf{500$\times$ faster} real-time execution 
while preserving near-optimal profit compared with MILP.
\begin{table}[htbp]
\centering
\caption{Computation time comparison}
\label{tab:runtime}
\begin{tabular}{lcc}
\toprule
\textbf{Method} & \textbf{Avg per step} & \textbf{Total time} \\
\midrule
Baseline (rule-based) & $\le$0.1\,ms & $\le$10\,s \\
Q-learning (training) & --- & 2.46\,h (offline) \\
Q-learning (inference) & 0.09\,ms & 9.52\,s \\
MILP (rolling horizon) & 50\,ms & 2.29\,h \\
\bottomrule
\end{tabular}
\end{table}

\subsubsection{Training diagnostic}
\begin{figure}[!htbp]
\setlength{\abovecaptionskip}{-0.1cm}  
    \setlength{\belowcaptionskip}{-0.1cm} 
    \centering
    \includegraphics[width=\columnwidth]{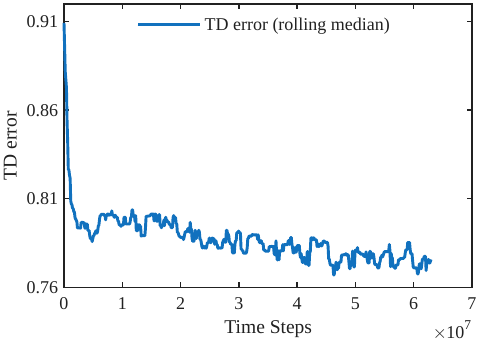}
    \caption{Value Function Convergence. TD error (rolling median, window $=9$) vs.\ time steps.
  Lower, flatter curves indicate smaller Bellman residuals and negligible $Q$ updates.}
    \label{fig:convergence}
\end{figure}

We monitor learning via the one-step temporal-difference (TD) error
$\delta_t = r_t + \gamma \max_{a'} Q(s_{t+1},a') - Q(s_t,a_t)$.
In tabular Q-learning the update magnitude equals $\alpha_t|\delta_t|$,
so a low and stable TD band implies small per-step changes of $Q$ on the
visited the state-action distribution and a stabilized greedy policy.
Fig.~\ref{fig:convergence} plots a rolling median of $|\delta_t|$ (window $=9$),
showing a rapid initial drop followed by a flat plateau, consistent with
stable value estimates.

\subsection{Sensitivity analysis}
\label{sec:sensitivity}

To further examine the robustness of the proposed Q-learning framework,
we perform sensitivity tests with respect to (i) the reward-smoothing factor $k$
and (ii) the furnace heterogeneity.

\subsubsection{Reward-smoothing factor $k$}

Table~\ref{tab:sensitivity_split} shows how the total profit varies with the smoothing parameter $k$.A smaller $k$ emphasizes instantaneous profit (aggressive operation), while a larger $k$ distributes rewards more evenly across the melt--tap cycle (conservative operation). When $k=13$, the learner achieves the highest profit of \$87{,}112, demonstrating that moderate reward smoothing provides the best trade-off between stability and reactivity to price dynamics.

\begin{table}[htbp]
    \centering
    \caption{Sensitivity of Q-Learning Profit to Reward Smoothing Factor $k$}
    \label{tab:sensitivity_split}
    \renewcommand{\arraystretch}{1.2}
    \begin{tabular}{cr  cr}
        \toprule

        $k$ & Profit (\$) & $k$ & Profit (\$) \\
        \cmidrule(r){1-2} \cmidrule(l){3-4} 
        20 & 81,406 & 15 & 77,831 \\
        19 & 80,641 & 14 & 80,504 \\
        18 & 81,174 & \textbf{13} & \textbf{87,112} \\ 
        17 & 81,217 & 12 & 60,227 \\
        16 & 77,849 & 11 & 0 \\
        \bottomrule
    \end{tabular}
\end{table}

\subsubsection{Heterogeneous unit configuration}

We further evaluate model robustness under heterogeneous furnace parameters
(High-efficiency, Standard, and Legacy) listed in Table~\ref{tab:paramsheter}.
As summarized in Table~\ref{tab:profit_summary_hetero},
Q-learning achieves 87{,}112~\$ profit, representing a 72.4\% uplift
over the baseline and reaching 89.8\% of the MILP-level profit.
This confirms that the proposed learning-based controller maintains
stable near-optimal performance even with differing unit dynamics and efficiencies.

\begin{table}[h]
\centering
\caption{Heterogeneous EAF unit parameters (Unit 1: high-efficiency, Unit 2: standard, Unit 3: legacy).}
\label{tab:paramsheter}
\setlength{\tabcolsep}{4pt} 
\renewcommand{\arraystretch}{1.2}
\begin{tabular}{lccccccccc}
\toprule
\multirow{2}{*}{\textbf{Furnace}} & \multirow{2}{*}{$I_i$} & \multirow{2}{*}{$R_i$} & \multirow{2}{*}{$K_i$} & \multirow{2}{*}{$\pi_i$} & \multirow{2}{*}{$\alpha_i$} & \multirow{2}{*}{$C_i$} & \multirow{2}{*}{$\delta_i$} & \multicolumn{2}{c}{Power [MW]} \\ 
\cmidrule(lr){9-10}
 & & & & & & & & $P^{\text{melt}}$ & $P^{\text{base}}$ \\ 
\midrule
Unit 1 & 1.2 & 1/10 & 1/15 & 420 & 0.88 & 280 & 80 & 0.0542 & 0.0029 \\
Unit 2        & 1.0 & 1/12 & 1/15 & 400 & 0.92 & 300 & 50 & 0.0367 & 0.0033 \\
Unit 3         & 0.8 & 1/15 & 1/20 & 380 & 0.92 & 300 & 20 & 0.0460 & 0.0042 \\
\bottomrule
\end{tabular}
\end{table}

\begin{table}[htbp!]
\centering
\caption{Profit over the 365-day horizon under heterogeneous unit configuration.}
\label{tab:profit_summary_hetero}
\vspace{1mm}
\setlength{\tabcolsep}{1.5pt}
\begin{tabular}{lrrr}
\toprule
Policy & \multicolumn{1}{c}{Cumulative profit [\$]} &
\multicolumn{1}{c}{Uplift vs Baseline} &
\multicolumn{1}{c}{Share of MILP} \\
\midrule
Baseline & 54{,}434 & --- & 67.0\% \\
Q-learning & 75{,}183 & +38.1\% & 92.5\% \\
MILP & 81{,}290 & +49.3\% & 100\% \\
\bottomrule
\end{tabular}
\end{table}

Overall, the sensitivity results show that the proposed Q-learning framework
generalizes well across both algorithmic (reward shaping) and physical
(unit heterogeneity) variations, consistently approaching the MILP benchmark
while retaining negligible online computation cost.

\subsection{Operational Insight: Flexibility under Power Constraints}

In the fixed-cycle scenarios, the system can activate at most two furnaces
due to the aggregate power limit $P^{\max}$.
This rigid scheduling often leads to idle periods where
a large, energy-intensive furnace remains partially active
at $P^{base}$ but cannot be turned off, resulting in wasted time and low utilization,
particularly during low-price hours.
Because startups are costly, such furnaces are reluctant to adjust frequently,
leading to inefficient use of available capacity.

The optimization-based methods, by contrast, exploit flexibility within the same
power envelope.
When the major furnaces operate below full load, 
the controller can opportunistically start smaller, more agile units
without exceeding $P^{\max}$.
These lightweight units capture low-price opportunities and increase overall throughput,
thus improving both profit and energy efficiency.

This insight suggests a practical design implication:
in real industrial systems, supplementing large, inflexible furnaces
with small and fast-response devices, such as electric heaters or hydrogen electrolyzers, 
can effectively fill low-load gaps, mitigate idle losses,
and enhance system-level profitability under grid power constraints.


\section{Discussion and Conclusion}\label{conclusion}

We proposed a mixed-integer linear programming optimization model to optimize the operation of EAF under volatile electricity prices. We formulated a rolling-horizon MILP that captures start-up costs, production delays, and shared feeder constraints, and paired it with a Q-learning controller that operates without commercial solvers and uses only day-ahead price signals. This combination allows us to benchmark economically optimal behavior under perfect foresight while evaluating a practical, real-time policy under non-anticipatory information.

Our results show three key findings. First, the MILP consistently aligns melting with low-price intervals and staggers multiple furnaces to respect feeder limits, achieving substantial gains relative to rule-based control. Second, our Q-learning dispatcher reliably reproduces these qualitative patterns and captures around 90\% of the MILP profit in both single- and multi-unit cases, confirming that EAF flexibility can be unlocked without solving large-scale optimization problems online. Third, our framework provides a tractable and scalable pathway for industrial deployment, bridging the gap between physics-based scheduling models and real-time control requirements.

Our work also highlights opportunities for future research. We simplified some thermochemical furnace dynamics and assumed perfect day-ahead forecasts in the MILP; incorporating process nonlinearities, uncertainty-aware optimization, and richer RL state representations would further enhance realism and robustness. Validating our approach with plant-level data and extending it to full melt-shop coordination represent natural next steps.

\bibliographystyle{IEEEtran}
\bibliography{references}

@inproceedings{pistorius_relating_2018,
	address = {Cham},
	title = {Relating {Reported} {Carbon} {Dioxide} {Emissions} to {Iron} and {Steelmaking} {Process} {Details}},
	isbn = {978-3-319-95022-8},
	doi = {10.1007/978-3-319-95022-8_49},
	language = {en},
	booktitle = {Extraction 2018},
	publisher = {Springer International Publishing},
	author = {Pistorius, P. Chris},
	editor = {Davis, Boyd R. and Moats, Michael S. and Wang, Shijie and Gregurek, Dean and Kapusta, Joël and Battle, Thomas P. and Schlesinger, Mark E. and Alvear Flores, Gerardo Raul and Jak, Evgueni and Goodall, Graeme and Free, Michael L. and Asselin, Edouard and Chagnes, Alexandre and Dreisinger, David and Jeffrey, Matthew and Lee, Jaeheon and Miller, Graeme and Petersen, Jochen and Ciminelli, Virginia S. T. and Xu, Qian and Molnar, Ronald and Adams, Jeff and Liu, Wenying and Verbaan, Niels and Goode, John and London, Ian M. and Azimi, Gisele and Forstner, Alex and Kappes, Ronel and Bhambhani, Tarun},
	year = {2018},
	pages = {631--638},
}

@article{stubbles_minimill_2009,
	title = {The {Minimill} {Story}},
	volume = {40},
	issn = {1073-5615, 1543-1916},
	url = {https://link.springer.com/10.1007/s11663-008-9216-9},
	doi = {10.1007/s11663-008-9216-9},
	language = {en},
	number = {2},
	urldate = {2025-12-07},
	journal = {Metallurgical and Materials Transactions B},
	author = {Stubbles, John R.},
	month = apr,
	year = {2009},
	pages = {134--144},
}

@misc{steel_alexandria_nodate,
	title = {Alexandria {Steelmaking} plant},
	url = {https://www.ezzsteel.com/ezz-steel-plants/alexandria-steelmaking-plant},
	language = {en},
	urldate = {2025-12-07},
	journal = {Ezz Steel},
	author = {Steel, Ezz},
}

@misc{noauthor_steelmaking_nodate,
	title = {Steelmaking at {North} {Star} {\textbar} {BlueScope}},
	url = {https://www.bluescope.com/our-steel/case-studies/Steelmaking-at-North-Star},
	language = {en},
	urldate = {2025-12-07},
}

@article{zhang_cost-effective_2017,
	title = {Cost-{Effective} {Scheduling} of {Steel} {Plants} {With} {Flexible} {EAFs}},
	volume = {8},
	copyright = {https://ieeexplore.ieee.org/Xplorehelp/downloads/license-information/IEEE.html},
	issn = {1949-3053, 1949-3061},
	url = {http://ieeexplore.ieee.org/document/7482812/},
	doi = {10.1109/TSG.2016.2575000},
	language = {en},
	number = {1},
	urldate = {2025-10-30},
	journal = {IEEE Transactions on Smart Grid},
	author = {Zhang, Xiao and Hug, Gabriela and Harjunkoski, Iiro},
	month = jan,
	year = {2017},
	pages = {239--249},
}

@article{li_constraint_2015,
	title = {Constraint {Programming} {Approach} to {Steelmaking}-making {Process} {Scheduling}},
	volume = {5},
	issn = {1941-6687},
	url = {https://scholarworks.lib.csusb.edu/ciima/vol5/iss3/2},
	doi = {10.58729/1941-6687.1267},
	language = {en},
	number = {3},
	urldate = {2025-10-30},
	journal = {Communications of the IIMA},
	author = {Li, Tieke and Li, Yan and Zhang, Qun and Gao, Xuedong and Dai, Shufen},
	month = jan,
	year = {2015},
}

@misc{noauthor_steel_nodate,
	title = {Steel melt shop is getting smarter},
	url = {https://new.abb.com/metals/digital-transformation-in-metals/smart-melt-shop/steel-melt-shop-getting-smarter?utm_source=chatgpt.com},
	language = {en},
	urldate = {2025-10-30},
	journal = {Metals},
}

@article{su_multi-objective_2023,
	title = {Multi-objective scheduling of a steelmaking plant integrated with renewable energy sources and energy storage systems: {Balancing} costs, emissions and make-span},
	volume = {428},
	issn = {0959-6526},
	shorttitle = {Multi-objective scheduling of a steelmaking plant integrated with renewable energy sources and energy storage systems},
	url = {https://www.sciencedirect.com/science/article/pii/S0959652623035084},
	doi = {10.1016/j.jclepro.2023.139350},
	urldate = {2025-10-30},
	journal = {Journal of Cleaner Production},
	author = {Su, Pengfei and Zhou, Yue and Wu, Jianzhong},
	month = nov,
	year = {2023},
	pages = {139350},
}

@article{abadi_review_2024,
	title = {A review of simulation and numerical modeling of electric arc furnace ({EAF}) and its processes},
	volume = {10},
	issn = {24058440},
	url = {https://linkinghub.elsevier.com/retrieve/pii/S240584402408188X},
	doi = {10.1016/j.heliyon.2024.e32157},
	language = {en},
	number = {11},
	urldate = {2025-10-30},
	journal = {Heliyon},
	author = {Abadi, Mahmoud Makki and Tang, Hongyan and Rashidi, Mohammad Mehdi},
	month = jun,
	year = {2024},
	pages = {e32157},
}

@article{zhao_two-stage_2024,
	title = {Two-stage day-ahead and intra-day scheduling considering electric arc furnace control and wind power modal decomposition},
	volume = {302},
	issn = {0360-5442},
	url = {https://www.sciencedirect.com/science/article/pii/S0360544224014671},
	doi = {10.1016/j.energy.2024.131694},
	urldate = {2025-10-30},
	journal = {Energy},
	author = {Zhao, Xudong and Wang, Yibo and Liu, Chuang and Cai, Guowei and Ge, Weichun and Wang, Bowen and Wang, Dongzhe and Shang, Jingru and Zhao, Yiru},
	month = sep,
	year = {2024},
	pages = {131694},
}

@article{saboohi_optimization_2019,
	title = {Optimization of the {Electric} {Arc} {Furnace} {Process}},
	volume = {66},
	issn = {1557-9948},
	url = {https://ieeexplore.ieee.org/abstract/document/8556389},
	doi = {10.1109/TIE.2018.2883247},
	number = {10},
	urldate = {2025-10-30},
	journal = {IEEE Transactions on Industrial Electronics},
	author = {Saboohi, Yadollah and Fathi, Amirhossein and Škrjanc, Igor and Logar, Vito},
	month = oct,
	year = {2019},
	pages = {8030--8039},
}

@article{lyu_efficient_2025,
	title = {Efficient {Scheduling} of {Discrete} {Industrial} {Processes} {Through} {Continuous} {Modeling}},
	volume = {16},
	issn = {1949-3061},
	url = {https://ieeexplore.ieee.org/document/11082423/authors},
	doi = {10.1109/TSG.2025.3589551},
	number = {6},
	urldate = {2025-10-30},
	journal = {IEEE Transactions on Smart Grid},
	author = {Lyu, Ruike and Su, Xiangbo and Du, Ershun and Guo, Hongye and Chen, Qixin and Kang, Chongqing},
	month = nov,
	year = {2025},
	pages = {4726--4740},
}

@article{hadera_optimization_2015,
	title = {Optimization of steel production scheduling with complex time-sensitive electricity cost},
	volume = {76},
	issn = {0098-1354},
	url = {https://www.sciencedirect.com/science/article/pii/S0098135415000472},
	doi = {10.1016/j.compchemeng.2015.02.004},
	urldate = {2025-10-30},
	journal = {Computers \& Chemical Engineering},
	author = {Hadera, Hubert and Harjunkoski, Iiro and Sand, Guido and Grossmann, Ignacio E. and Engell, Sebastian},
	month = may,
	year = {2015},
	pages = {117--136},
}

@article{mays_quasi-stochastic_2021,
	title = {Quasi-{Stochastic} {Electricity} {Markets}},
	volume = {3},
	issn = {2575-1484, 2575-1492},
	url = {https://pubsonline.informs.org/doi/10.1287/ijoo.2021.0051},
	doi = {10.1287/ijoo.2021.0051},
	language = {en},
	number = {4},
	urldate = {2025-10-21},
	journal = {INFORMS Journal on Optimization},
	author = {Mays, Jacob},
	month = oct,
	year = {2021},
	pages = {350--372},
}

@article{li_socially_2024,
	title = {Socially optimal energy usage via adaptive pricing},
	volume = {235},
	issn = {03787796},
	url = {https://linkinghub.elsevier.com/retrieve/pii/S0378779624005261},
	doi = {10.1016/j.epsr.2024.110640},
	language = {en},
	urldate = {2025-10-21},
	journal = {Electric Power Systems Research},
	author = {Li, Jiayi and Motoki, Matthew and Zhang, Baosen},
	month = oct,
	year = {2024},
	pages = {110640},
}

@article{qi_locational_2025,
	title = {Locational {Energy} {Storage} {Bid} {Bounds} for {Facilitating} {Social} {Welfare} {Convergence}},
	issn = {2771-9626},
	url = {https://ieeexplore.ieee.org/abstract/document/11034735},
	doi = {10.1109/TEMPR.2025.3579671},
	urldate = {2025-09-18},
	journal = {Policy and Regulation IEEE Transactions on Energy Markets},
	author = {Qi, Ning and Xu, Bolun},
	year = {2025},
	pages = {1--12},
}

@article{zheng_energy_2023,
	title = {Energy {Storage} {State}-of-{Charge} {Market} {Model}},
	volume = {1},
	issn = {2771-9626},
	url = {https://ieeexplore.ieee.org/document/10021874/},
	doi = {10.1109/TEMPR.2023.3238135},
	number = {1},
	urldate = {2025-09-18},
	journal = {Policy and Regulation IEEE Transactions on Energy Markets},
	author = {Zheng, Ningkun and Qin, Xin and Wu, Di and Murtaugh, Gabe and Xu, Bolun},
	month = mar,
	year = {2023},
	pages = {11--22},
}

@article{lee_optimization_2023,
	title = {Optimization of iron and steel manufacturing plant considering electricity price tariff and electric arc furnace control},
	volume = {17},
	issn = {1751-8687, 1751-8695},
	url = {https://ietresearch.onlinelibrary.wiley.com/doi/10.1049/gtd2.13017},
	doi = {10.1049/gtd2.13017},
	language = {en},
	number = {22},
	urldate = {2025-09-08},
	journal = {IET Generation, Transmission \& Distribution},
	author = {Lee, Seok‐Young and Lee, Gyu‐Sub and Moon, Seung‐Il and Yoon, Yong‐Tae},
	month = nov,
	year = {2023},
	pages = {5027--5040},
}

@article{fan_low-carbon_2021,
	title = {Low-carbon production of iron and steel: {Technology} options, economic assessment, and policy},
	volume = {5},
	issn = {2542-4351},
	shorttitle = {Low-carbon production of iron and steel},
	url = {https://www.sciencedirect.com/science/article/pii/S2542435121000957},
	doi = {10.1016/j.joule.2021.02.018},
	number = {4},
	urldate = {2025-09-04},
	journal = {Joule},
	author = {Fan, Zhiyuan and Friedmann, S. Julio},
	month = apr,
	year = {2021},
	pages = {829--862},
}

@article{kim_decarbonizing_2022,
	title = {Decarbonizing the iron and steel industry: {A} systematic review of sociotechnical systems, technological innovations, and policy options},
	volume = {89},
	issn = {2214-6296},
	url = {https://www.sciencedirect.com/science/article/pii/S2214629622000706},
	doi = {https://doi.org/10.1016/j.erss.2022.102565},
	journal = {Energy Research \& Social Science},
	author = {Kim, Jinsoo and Sovacool, Benjamin K. and Bazilian, Morgan and Griffiths, Steve and Lee, Junghwan and Yang, Minyoung and Lee, Jordy},
	year = {2022},
	pages = {102565},
}

@article{xu_optimal_2018,
	title = {Optimal {Battery} {Participation} in {Frequency} {Regulation} {Markets}},
	volume = {33},
	issn = {1558-0679},
	url = {https://ieeexplore.ieee.org/abstract/document/8383984},
	doi = {10.1109/TPWRS.2018.2846774},
	number = {6},
	urldate = {2025-08-26},
	journal = {IEEE Transactions on Power Systems},
	author = {Xu, Bolun and Shi, Yuanyuan and Kirschen, Daniel S. and Zhang, Baosen},
	month = nov,
	year = {2018},
	pages = {6715--6725},
}

\end{document}